# Low-complexity full-field ultrafast nonlinear dynamics prediction by a convolutional feature separation modeling method

HANG YANG, HAOCHEN ZHAO, ZEKUN NIU, GUOQING PU, SHILIN XIAO, WEISHENG HU, AND LILIN YI[*]

*State Key Lab of Advanced Optical Communication Systems and Networks, School of Electronic Information and Electrical Engineering, Shanghai Jiao Tong University, Shanghai 200240, China.*
*Corresponding author: lilinyi@sjtu.edu.cn*

**Abstract:** The modeling and prediction of the ultrafast nonlinear dynamics in the optical fiber are essential for the studies of laser design, experimental optimization, and other fundamental applications. The traditional propagation modeling method based on the nonlinear Schrödinger equation (NLSE) has long been regarded as extremely time-consuming, especially for designing and optimizing experiments. The recurrent neural network (RNN) has been implemented as an accurate intensity prediction tool with reduced complexity and good generalization capability. However, the complexity of long grid input points and the flexibility of neural network structure should be further optimized for broader applications. Here, we propose a convolutional feature separation modeling method to predict full-field ultrafast nonlinear dynamics with low complexity and high flexibility, where the linear effects are firstly modeled by NLSE-derived methods, then a convolutional deep learning method is implemented for nonlinearity modeling. With this method, the temporal relevance of nonlinear effects is substantially shortened, and the parameters and scale of neural networks can be greatly reduced. The running time achieves a 94% reduction versus NLSE and an 87% reduction versus RNN without accuracy deterioration. In addition, the input pulse conditions, including grid point numbers, durations, peak powers, and propagation distance, can be flexibly changed during the predicting process. The results represent a remarkable improvement in the ultrafast nonlinear dynamics prediction and this work also provides novel perspectives of the feature separation modeling method for quickly and flexibly studying the nonlinear characteristics in other fields.

## 1. Introduction

Most systems inherently exhibit some interactions between linearities and nonlinearities, and system outputs always change in variables over time [1, 2]. The complex nonlinear dynamics have long been regarded as challenging and unpredictable in many research fields, including biology, chemistry, hydrodynamics, and engineering [3-5]. In terms of optics and photonics, nonlinear pulse propagation in optical fiber waveguides is a typical complex nonlinear evolution, especially for high-power ultrafast pulses. The analysis, control, and prediction of the ultrafast nonlinear dynamics in optical fiber are of great importance in the development of laser design, experimental optimization, remote sensing, and other fundamental researches [6-10]. The nonlinear Schrödinger equation (NLSE) can accurately describe the pulse propagation in the optical fiber, and the split-step Fourier method (SSFM)-based numerical solution has been proven with high accuracy compared with pulse transmission in the experiment [11, 12]. However, the numerical methods typically involve high computation costs due to the requirement of many iterative complex operations. The longtime simulation imposes a severe bottleneck in using conventional methods to predict the complete pulse propagation in the optical fiber [13].

Recently, machine learning has been used for nonlinear dynamics prediction or fiber transmission modeling due to its strong nonlinear data fitting ability, flexible design structure, and low computation complexity. The physical-informed neural network (PINN) was proposed to study the propagation of pulses based on NLSE [14], and the nonlinear dynamics of the pulses can be predicted by applying physical knowledge under some conditions [15-16]. However, when the initial pulse parameters change, the PINN should be re-trained. Namely, the generalization ability of the PINN-based method is extremely limited. The recurrent neural network (RNN) has been regarded as a good prediction tool for time-series data and applied to predict ultrafast nonlinear dynamics with high accuracy and good generalization for multiple conditions [17]. This work aims to predict the intensity profiles of the time domain and spectral domain by two separate NNs. And the phase information of pulses is neglected. In addition, due to the internal loop unit of RNN, the parallelism and complexity of RNN are limited. A feed-forward neural network (FNN) is introduced to predict intensity and phase profiles together in the temporal (or spectral) domain with lower complexity but higher prediction errors [18]. In the work of RNN and FNN, the input pulses are required to be truncated or down-sampled to keep at short input grid points and low complexity [17, 18]. The comparison to the NLSE simulation with complete long grid points seems not relatively fair. When the input grid points cannot be turned to short length manually in some cases and are as long as NLSE simulation, the computational complexity of RNN and FNN will be much higher. Furthermore, these structures can only predict the input pulse with the fixed input length, so they can not realize the prediction of the flexible pulse input length. Therefore, a faster and more flexible prediction method for full-field ultrafast nonlinear dynamics with long input grid points is still an open issue.

Here, we focus on the full-field pulse propagation modeling by collecting full-length temporal pulses represented by complex numbers as training datasets. It inherently includes the phase information by phase calculation from complex numbers and the spectral information by performing the Fourier transform over the temporal pulses. This eliminates the need for two separate models for temporal and spectral domains, and the complex number representation allows applying some physical knowledge to the pulses. For fast and accurate ultrafast dynamics predictions, the linear features are modeled by NLSE-derived methods, and nonlinear features are modeled by a dedicatedly-designed convolutional neural network (CNN). This convolutional feature separation modeling (FSM) method has advantages as follows (due to the accuracy degradation of the FNN, our work is mainly compared with the RNN):

(1) The temporal relevance of nonlinear effects between the grid points can be reduced by the FSM, resulting in a short local correlation in the time axis. In this case, the CNN structure with much shorter kernel lengths can achieve accurate nonlinearity modeling. The required parameters of CNN are only 0.14% of RNN at the 2048 pulse grid points. The running time of NLSE, RNN, and CNN at 2048 points are 763s, 46s, and 5.9s, which demonstrates a 94% reduction versus NLSE and an 87% reduction versus RNN.

(2) Compared with RNN, the simple CNN structure with FSM can achieve equal accuracy within the training transmission distances and higher accuracy beyond the training distances. This local correlation calculation structure is demonstrated with high stability and good distance generalization capability.

(3) Based on the convolution computation, pulse propagation with dynamic input pulse lengths can be flexibly achieved by CNN. On one hand, the pulse can be truncated to a shorter length for complexity reduction. On the other hand, the input length can be elongated for multiple pulse propagation.

We show the generalization ability for different pulse durations, peak powers, grid points, and transmission distances beyond the training conditions. The results demonstrate that the CNN with FSM is a full-field ultrafast nonlinear dynamic prediction method with low complexity and high flexibility. This work also provides a new perspective of feature separation modeling with local correlation calculations for fast and accurate nonlinear dynamics studies in other fields.

## 2. Principle of the convolutional feature separation modeling method

Ultrafast pulse propagation in optical fiber has rapid nonlinear changes determined by the interplay between a range of nonlinear and linear effects. Specifically, the linear effects always lead to long temporal dependency between the grid points of pulses [19]. The nonlinear effects are determined by the input pulse power and the fiber channel characteristics. The interaction between the linear and nonlinear features accumulates along with transmission. To predict nonlinear dynamics fast and accurately, the proposed prediction method considers the temporal and distance dependency of pulses by separately modeling the linear and nonlinear features and injecting the multiple preceding known pulses to NN.

We use the numerical solution method to generate training data at different distances with the same distance interval. Note that the distance interval of the training dataset is set larger to reduce the complexity compared to that in the NLSE simulation. The data is represented by the real and imaginary parts, so the temporal intensity profiles and phase profiles can be directly generated by one NN. And the spectral profiles can also be easily obtained by Fast Fourier Transform (FFT).

The schematic of the convolutional FSM architecture is presented in Fig. 1a, where the model is divided into the NLSE-derived linear model and the CNN-based nonlinear model. The CNN model contains the input layer, 1-dimensional convolution (conv1d) layers [20], dense layers, and output layer. The iterations are implemented for the longer transmission distance.

A detailed scheme of the CNN structure with FSM and the data arrangement are illustrated in Fig. 1b. First, the pulses from previous distances (form $z$-10$\Delta z$ to $z$-$\Delta z$) should be collected while predicting the output pulse of distance z. $\Delta z$ is the interval distance we set along the propagation direction. When predicting the first interval distance ($z = \Delta z$), the input previous pulses are the same as the input pulses, i.e. $A(-9\Delta z), …, A(-\Delta z) = A(0)$, where $A$ represents the optical pulse profile. The input pulses of multiple distances can improve the prediction accuracy and the optimal distances number is set to 10 by the ergodic parameters search. During the prediction process, the output pulse of distance z is fed back to the input for the pulse prediction of the next distance $A(z+\Delta z)$. The iteration operation also provides NN with the distance generalization ability.

Second, the linear feature modeling (LFM) is implemented in one interval distance by prior knowledge, which can be expressed by

$$\tilde{A}(\omega, z + \Delta z) = \tilde{A}(\omega, z) \exp\left( j \left( \sum_{k \geq 2} \frac{\beta_k}{k!} w^k \right) \Delta z \right), \tag{1}$$

where $\omega$ is the frequency component, $\beta_k$ is the dispersion coefficients associated with the Taylor series expansion of the propagation constant $\beta(\omega)$. This linear modeling operation can achieve the long-time dependency feature modeling in one step. This process is based on the physics knowledge described by NLSE. Therefore, the linear features can be modeled accurately and not depend on the neural network performances.

After the linear modeling operation, conv1d layers and local dense layers are used for nonlinear feature modeling. The convolution kernel length and dense layers dimension we use is relatively small, where a hypothesis that the nonlinear features are short-time dependency is introduced. If linear and nonlinear are modeled together, NN is required to model the global long-time correlation, which will lead to a rapid increase in complexity and the number of parameters. We adopt the method of linear and nonlinear FSM, which can realize local correlation modeling and effectively reduce the parameters and complexity of NN.

As shown in Fig. 1b, the detailed CNN structure for nonlinear feature modeling consists of convolutional layers and fully connected layers with shared parameters. The input data is arranged in the form of a two-dimensional matrix, and its dimension is expressed as ($D\times2P$), where *D* represents the pulses number of preceding distances, and *P* represents the complex grid points of a pulse. The 2*P* term refers to real (*R*) and imaginary (*I*) parts of *P* grid points, because the inputs of NN are typically real numbers. The use of a complex-valued neural network is of future discussion and is not covered. After rearrangement, several layers of one-dimensional convolution operation are carried out for local correlation extraction. Each convolution operation contains multiple channels (*Ch*). The size of the convolution kernel relates to the output dimension we designed. We keep the output dimension of each convolutional layer at (*Ch*×*P*), which can be achieved by setting the stride size and padding number.

After obtaining the convolutional layer output, the data of each column (dimension is *Ch*) is processed by a small fully-connected NN respectively and mapped to the prediction data corresponding to the time grid point. Compared to the method of expanding into one-dimensional data and then fully connecting, our proposed method can effectively reduce the scale and complexity of the dense NN structures. After the local correlation extraction by the convolutional layer, the output data contain the coincident distribution or characteristics. On the basis of this intuition, we share the parameters among the fully-connected NNs, which enables the NN structure with fewer parameters and fast convergence. The input of each dense layer is (*Ch*×1) and the dense layer output dimension is set as 2, representing a complex point. The total number of the fully-connected NNs is *P*, corresponding to the column number of convolutional layer output. The outputs of these fully-connected NNs are concatenated to form the output with length 2*P*, which represents the pulse prediction of the next distance *z*.

During the training process, the mean absolute error (MAE) loss between the predicted pulse and the actual pulse in the label is calculated for NN training. Through the hyperparametric search, the number of convolution layers we set is 3 and the fully connected hidden layer is 4. The channel number of the kernel is 90, i.e. $Ch = 90$. We select the convolution kernel size as 18 for the first conv1d layer, and 9 for the other two conv1d layers. If the FSM method is not used, the kernel size should be larger to maintain high prediction accuracy. The hidden dimension of the dense layer is the same as the input layer. The optimizer is RMSprop [21] and the epoch number is 120. The learning rate is initialized as 1e-3 and then decreases with the epochs based on a cosine annealing with the warm restart schedule [22].

The above descriptions stress that CNN with FSM can predict the full-field ultrafast nonlinear dynamics without any pulse truncation by one NN. Meanwhile, the convolution kernel is fed with fixed-length data at a time and then slid continuously to cover the full input data length. Therefore, one convolution kernel flexibly supports the dynamic processing of different input pulse lengths, which allows truncated or extended pulses beyond the points number of the training pulses. We demonstrate two applications of dynamic input pulse lengths capability in predicting phase. The first one, as shown in Fig. 2, is to dynamically truncate pulses to only apply the NN model on the high-energy portion. This can help reduce prediction calculation costs, which will be further discussed in the complexity section. The second one is to accept multiple input pulses in one elongated window. This is useful to apply a trained model on longer input pulse lengths without re-training, which will be further investigated in the generalization section.

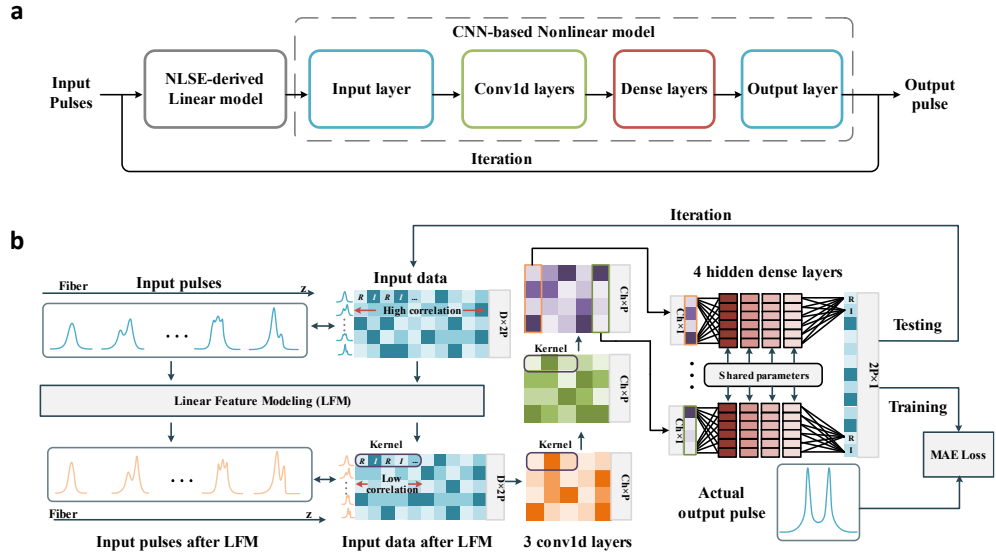

Fig. 1 The architecture of the convolutional neural network (CNN) with feature separation modeling (FSM). **a,** Schematic of the convolutional FSM architecture, showing the NLSE-derived linear model, CNN-based nonlinear model and the iteration steps for longer propagation. **b,** The detailed structure and data arrangement of the CNN with FSM. The pulses from the previous distances are implemented by linear feature modeling (LFM) first. And the pulse data are arranged to form a two-dimensional matrix D $\times$ 2P, where D is 10 presenting the previous distances number, P is the pulse grid points, and 2P represents each complex number represented by two real numbers. Then 3 conv1d layers and 4 hidden dense layers with shared parameters are applied for the nonlinear feature modeling. The Ch represents the channel number of the kernel in each conv1d layer. The matrices are shown by rectangles with different colors representing different processing stages, and the color depth of each small square in the matrix indicates the data values.

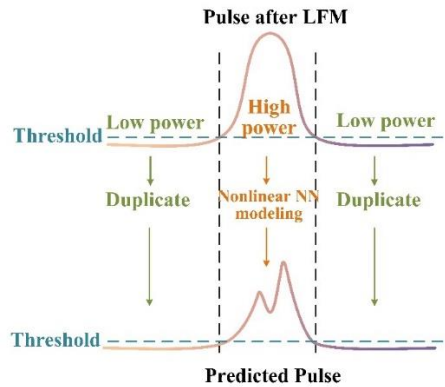

Fig. 2 The diagram of the dynamic input pulse length to CNN based on FSM. A threshold can be defined by the user demands and channel conditions in each step to separate the high power and low power pulses. The pulses after LFM with high power are input to the CNN for nonlinear modeling. The pulses after LFM with low power are directly duplicated to the output predicted pulse.

## 3. The nonlinear dynamics modeling setup and results

### 3.1 The modeling setup

To verify the nonlinear dynamics prediction capability of the proposed CNN structure based on the hybrid modeling method, we need to obtain the fully nonlinear evolution map of injected ultrafast pulses in the optical fiber as the training data set. We chose two typical ultrafast nonlinear scenarios: higher-order soliton (HOS) compression, and broadband optical super-continuum (SC) generation [23]. This FSM method and CNN structure can also be applied to other NLSE-like cases. The parameters for both cases are selected by the previous typical works and listed in Table 1 [12, 23]. In the first case, the pulse duration $\Delta\tau$ (full-width at half-maximum, FWHM) and peak power $P_0$ are randomly varied from 0.77-1.43ps and 18.4-34.2W. The soliton number $N = \sqrt{\gamma P_0 T_0^2/|\beta_2|}$ in this field is varied from 3.5 to 8.9 ($\gamma$ is the nonlinear parameter, $\beta_2$ is the group velocity dispersion parameter, and $T_0 = \Delta\tau/1.763$). The total transmission distance of the training data is 1300cm and the step distance length of NN is 13cm, so the total iteration steps of NN is 100. We demonstrate the input pulse points number, duration, peak power, and distance generalization beyond the training dataset conditions in this case. The input peak power of SC generation varied from 500W to 2000W, and the pulse duration is fixed at 0.1ps. SC generation process includes more complex nonlinear dynamics induced by Raman and self-steepening effects. The results of the SC generation verified the ability of complex nonlinear dynamic prediction. The grid points of the training pulse are 1024 and 2048 for HOS compression and SC generation respectively. We emphasize that the full pulse points can be predicted accurately and fast by the proposed CNN structure with FSM.

**Table 1. Modeling parameters**

| Scenarios | HOS compression | | SC generation | |
|---|---|---|---|---|
| FWHM (ps) | [0.77,1.43] | | 0.1 | |
| Input peak power (W) | [18.41,34.19] | | [500,2000] | |
| Fiber length (cm) | 1300 | | 20 | |
| Step length of NN (cm) | 13 | | 0.1 | |
| Step length of NLSE (cm) | 0.13 | | 0.001 | |
| Temporal window size (ps) | 10 | | 5 | |
| Grid points | 1024 | | 2048 | |
| Training sets | 2900 | | 1250 | |
| Testing sets | 100 | | 50 | |
| Center wavelength (nm) | 1550 | | 810 | |
| Dispersion coefficient (ps$^{order}$/km) | $\beta_2, \beta_3$ | -5.23, 4.27e-2 | $\beta_2, \beta_3, \beta_4, \beta_5, \beta_6, \beta_7$ | -9.59, 0.0784, -6.84e-05, -4.78e-07, 2.71e-09, -5e-12 |
| Nonlinear parameter (W$^{-1}$m$^{-1}$) | 18.4e-3 | | 0.1 | |
| Fiber channel effects | Self-phase modulation | | Self-phase modulation, Raman, Self-steepening | |

**3.2 Results**
**Comparisons of different methods**

To express the prediction gap between the NN-based model and the NLSE simulation quantitatively, we record the prediction errors of different cases and conditions. The root normalized mean square error (RNMSE) can be described as

$$RNMSE = \frac{1}{n}\sum_n \sqrt{\frac{\sum_{p,d}\|A_{n,p,d}-\hat{A}_{n,p,d}\|^2}{\sum_{p,d}\|A_{n,p,d}\|^2}}, \quad (2)$$

where $A$ and $\hat{A}$ represents the transmitted pulse represented by complex values simulated by NLSE and NN-based model. The variables $p$ and $d$ represent the grid point and distance index. $n$ denotes the number of random input pulse conditions.

First, the RNN structure and our designed CNN structure with and without FSM are compared. The RNN structure we test is the same as that in [17], consisting of 1 recurrent layer, 2 hidden dense layers, and the output layer. The output dimension of the recurrent layer is the same as that of the input and the previous pulse window is 10. Each input pulse consists of 1024 grid points in the HOS compression scene, so the input dimension of each RNN cell is 1024 for full-length pulse prediction. The kernel size of CNN is set to 18 whether FSM is performed or not, which leads to a fair prediction accuracy comparison under the same complexity.

As shown in Fig. 3, the RNMSE vs. distance is presented to show the iteration errors during the transmission. The results are tested under the conditions of HOS compression. Each RNMSE is computed over the $n = 100$ full-time pulses in the testing dataset with different input conditions. The distance before 13m is trained, and the distance after 13m is not trained to test generalization. For both methods, prediction error accumulates over the propagation distances, because the prediction errors at the previous distances will accumulate and further reduce the accuracy at the subsequent distances. The RNMSE of CNN at 13m showed a significant improvement from 0.15 without FSM to 0.09 with FSM. This highlights the effectiveness of FSM for CNN with small kernel sizes. Other simulations are also implemented to show that the same level of accuracy can be achieved without FSM by increasing the convolution kernel size (larger than 82), but the complexity will significantly increase and the model with many parameters will be hard to train. As for RNN, FSM operation has no effect on the prediction accuracy in this full-length input structure. Since the full-length grid points are input to the RNN cell, global points computation can be achieved, and thus realize good capability for channel effects (linear and nonlinear effects together) modeling. With the FSM, the input points of each RNN cell can be shortened by a sliding length-limited window for full-length modeling, which can reduce the input grid point number and the computational complexity. Compared with the CNN structure, this operation of RNN is required to be designed manually, resulting in a new data processing flow, which is not within the scope of our article. Note that the errors of RNN after 13m increase sharply, which indicates the weakness of distance generalization. On the contrary, the iterative errors of CNN increase steadily and keep at a lower level than RNN, which demonstrates the good distance generalization ability and stability of the local corrections calculation structure. We emphasize that while ensuring high performance, CNN with FSM has much lower complexity, which will be further described in the complexity section.

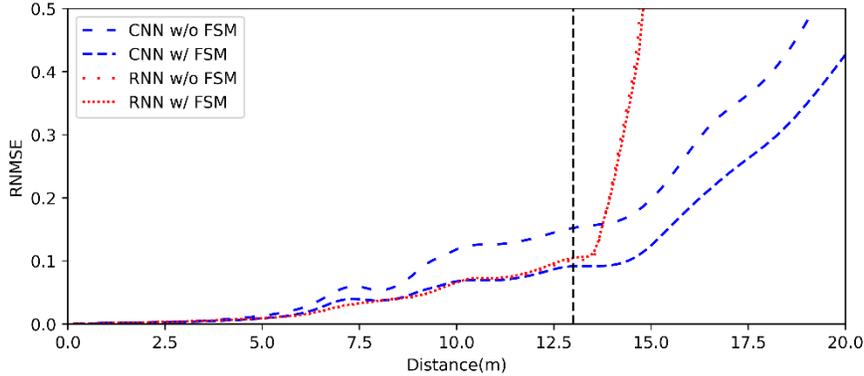

Fig. 3 The RNMSE vs. distances of CNN and RNN with/without FSM. The RNMSE at each distance is averaged over the 100 full-time pulses in the testing dataset with random input conditions.

**Accuracy**

We also present the detailed results of HOS compression and SC generation to present the prediction accuracy of CNN with FSM. The overall pulse evolution with transmission distance and detailed pulse profile at some distances are illustrated. The results are simulated using NLSE and predicted by CNN with FSM for accuracy comparison.

Fig. 4 illustrates the temporal intensity evolution (Fig. 4a), spectral intensity evolution (Fig. 4b), and phase changes varied with time and distance (Fig. 4c) of HOS propagation dynamics. The input peak power of these results is 30W and the duration is 0.8ps. The corresponding soliton number is about 4.7. These conditions are never included in the training dataset. From Fig. 4a and Fig. 4b, one can see the high consistency between the pulse propagation predicted by CNN with FSM and that simulated from NLSE on both temporal intensity evolution and spectral intensity evolution. Note that the narrowest pulse presented in temporal intensity evolution and maximum expansion presented in spectra intensity evolution have an agreement in distance value, which demonstrates the prediction accuracy of maximal compression distance. For clearer visual comparison, we select three distances (1.3m, 7.8m, and 10.4m) to plot the intensity profile in the temporal and spectra domain, where a great deal of overlap of full-time and full-spectral intensity pulse are shown. The RNMSE over the full temporal evolution is calculated as 0.028, which is a lower value than that of RNN and indicates the high accuracy of the pulse prediction during the full evolution.

We also plot the pulse phase profiles in the time domain with different distances in Fig. 4c to demonstrate the phase prediction accuracy of CNN with FSM. For the *pth* grid point represented by a complex number, the phase can be calculated as $Phase(p) = arctan\left[\frac{I(p)}{R(p)}\right]$, where *I* and *R* represent imaginary and real part respectively. The phase value is in the range of [-π, π], which can result in a discontinuous phase curve. Therefore, the phase unwrapping algorithm is used to present a continuous and actual phase cures shape [24]. It can be seen that the phase curves of NLSE simulation and CNN prediction have great coincidence, demonstrating high phase prediction accuracy. The full field nonlinear dynamics prediction is successfully achieved by the proposed CNN structure with FSM.

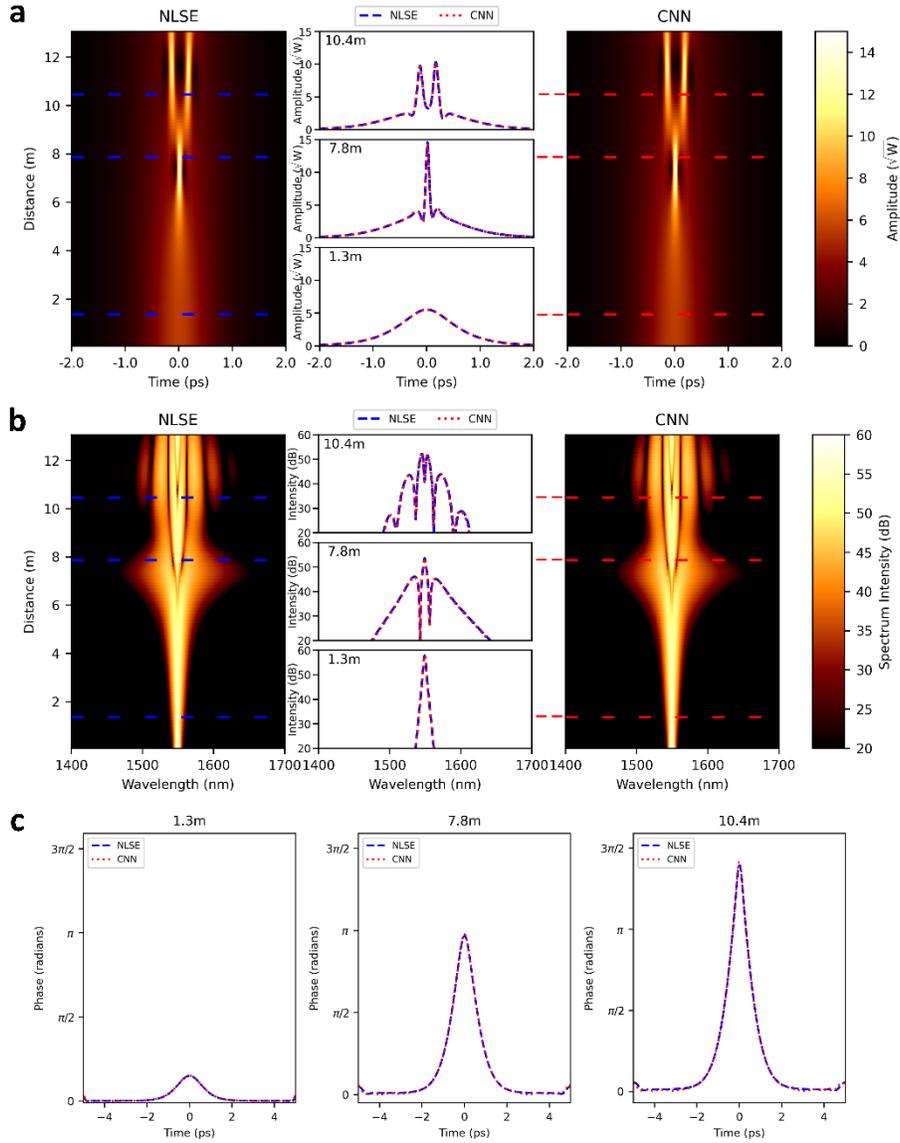

Fig. 4 Temporal evolution, spectra evolution and pulse profile of high-order soliton compression simulated by NLSE and predicted by CNN with FSM. **a-b**, Temporal (a), and spectral (b) intensity evolution of NLSE (left panel) and CNN (right panel), and comparison between the predicted (red lines) and simulated (blue lines) profiles at selected distances (middle panel). **c**, Pulse phase profiles in time domain simulated by NLSE (blue dash lines) and predicted by CNN (red dotted lines). All the results correspond to the input pulse with 0.8ps duration and 30W input peak power. The corresponding soliton number is about 4.7.

We next implement CNN with FSM to predict more complex nonlinear dynamics, i.e. the generation of a broadband SC. This case considers the femtosecond pulses propagations in the highly nonlinear fiber with anomalous dispersion, where the delayed Raman response and self-steepening effects produce stronger nonlinear dynamics than self-phase modulation [25, 26]. Results are shown in Fig. 5a, and Fig. 5b for temporal and spectral intensity evolution respectively. The associated intensity profiles at 1cm, 6cm, and 16cm are also plotted. The input peak power of these results is 1.6 kW. The excellent visual agreement between simulated and predicted evolution maps can be seen from the temporal and spectral domains. For temporal evolution, the generation process induced by high-order dispersion and Raman perturbation, including initial compression, soliton fission, and dispersive emission [23], can be perfectly described by the CNN. For spectral evolution, the generation processes, including dispersive wave generation, continuous redshift, and extreme broadband spectrally output, are perfectly reproduced by CNN. The RNMSE over the full spectral evolution is 0.08, which is a little higher than that of the HOS compression case since the more complex nonlinear dynamics, but it is still at a high accuracy level demonstrated by the accurate pulse profiles. Fig. 5c plot the pulse phase profiles in spectral domain simulated by NLSE and predicted by CNN. Since the drastic phase change with the frequency, the phase unwrapping algorithm is also used for clear phase curve comparations. The high coincidence proves the phase prediction accuracy by CNN with FSM.

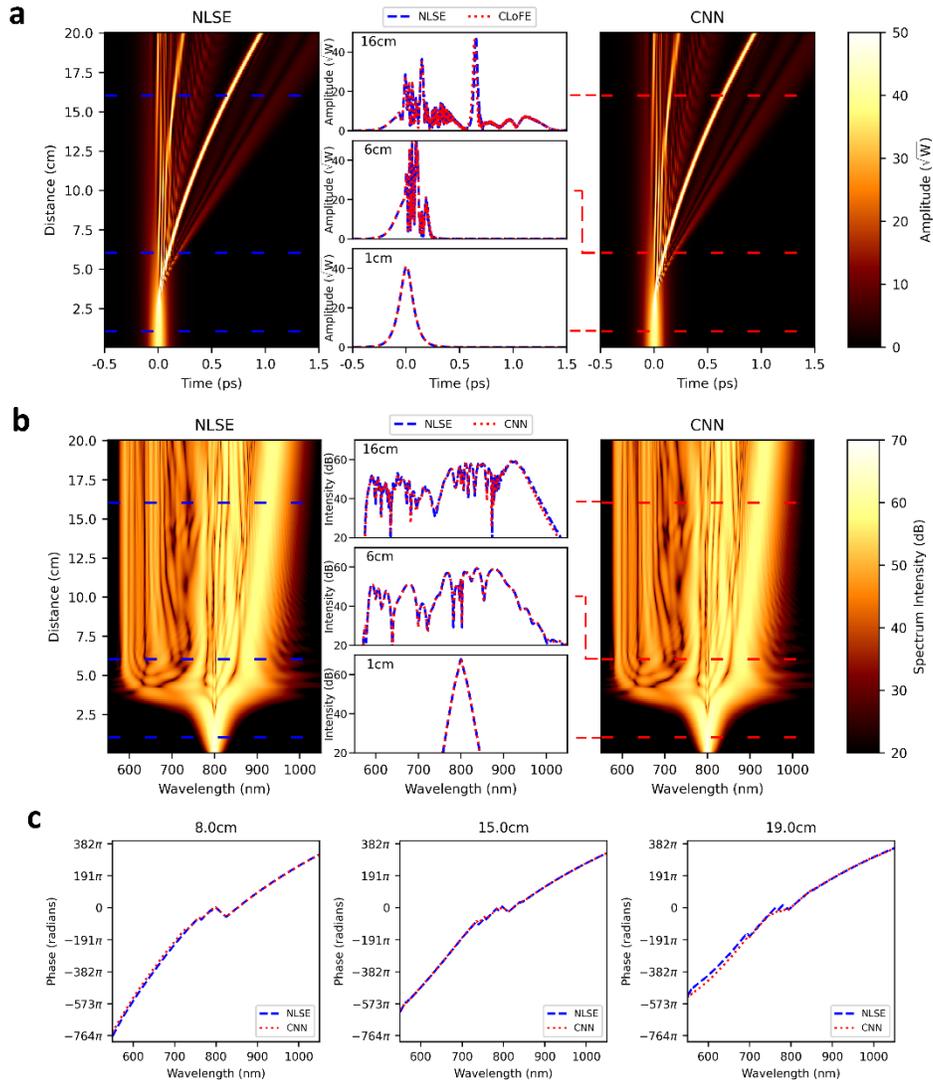

Fig. 5 Temporal evolution, spectra evolution and pulse profile of supercontinuum generation simulated by NLSE and predicted by CNN with FSM. **a-b**, Temporal (a) and spectral (b) intensity evolution of NLSE (left panel) and CNN (right panel), and comparison between the predicted (red lines) and simulated (blue lines) profiles at selected distances (middle panel). **c**, Pulse phase profiles in spectral domain simulated by NLSE (blue dash lines) and predicted by CNN (red dotted lines). All the results correspond to the input pulse with 0.1ps duration and 1.6kW input peak power.

**Generalization**
It is well known that the generalization capability of a NN is of huge challenge and an important criterion for applications. Only if a neural network can effectively generalize the input conditions it can have a wide range of applications in practice. For ultrafast pulse propagation, different input conditions always lead to different nonlinear evolution maps.

The input peak power and pulse duration are usually the points of interest for ultrafast pulse studies. The generalization for durations and peak power was reported in previous literature, where RNN is investigated [17]. Here, we demonstrate the generation capability of CNN for different input conditions. The durations and peak power are randomly varied from 0.77-1.43ps and 18.4-34.2W in the training dataset. Therefore, the designed CNN can realize accurate predictions among these ranges. To demonstrate the generalization ability of CNN with FSM, we show the temporal intensity profile with different durations and input peak powers in the HOS compression case. These selected conditions are handed picked and simulated separately. We investigate the generalization against input pulse conditions in the upper three rows of Fig. 6. In Fig. 6a and Fig. 6b, the durations are both 0.8ps, and the input peak powers are 20W and 30W respectively. In Fig. 6a and Fig. 6c, input peak powers are both 20W, and the durations are 0.8ps and 1.4ps. Under these three sets of conditions, the input pulses compressed and fissured at different rates, causing vastly different pulses shapes at the same distance. Pulses simulated by NLSE and CNN show a great deal of overlap at distances, which shows the generalization ability against input pulses durations and peak powers.

Besides generalization over input conditions, we next test and demonstrate the ability of CNN to predict distances farther than the training set, which is the first investigation for the distance generalization capability. The fourth row of Fig. 6 shows the prediction result at 19.6m, 150% of the training transmission length of 13m. The results are acquired by further iterating the CNN model by another 50 steps beyond the original 100 steps. Although the RNMSE of the pulse at 19.6m is larger than that within the training distance conditions, the temporal peak location and intensity agree very well between NLSE and CNN prediction. The pulse difference mainly comes from the low energy part of the pulse base, and the pulse at high power is accurate to maintain the pulse shape and reflect the nonlinear dynamics. The ability to predict beyond the distances in training data shows that CNN does learn the nonlinearity features of ultrafast pulse propagation. As far as we know, such generalization has not yet been reported in the literature concerning ultrafast pulse propagation modeling.

The flexibility of CNN also enables modeling multiple input pulses in one elongated window. In Fig. 7 we extended the temporal window size from 10ps to 40ps and corresponding grid points from 1024 to 4096. This enabled inputting 4 pulses of different durations and peak powers at once. All peaks propagated with great accuracy compared to NLSE simulations. This example demonstrated CNN's generalization ability of pulse duration, power, transmission distance, and simulation grid size.

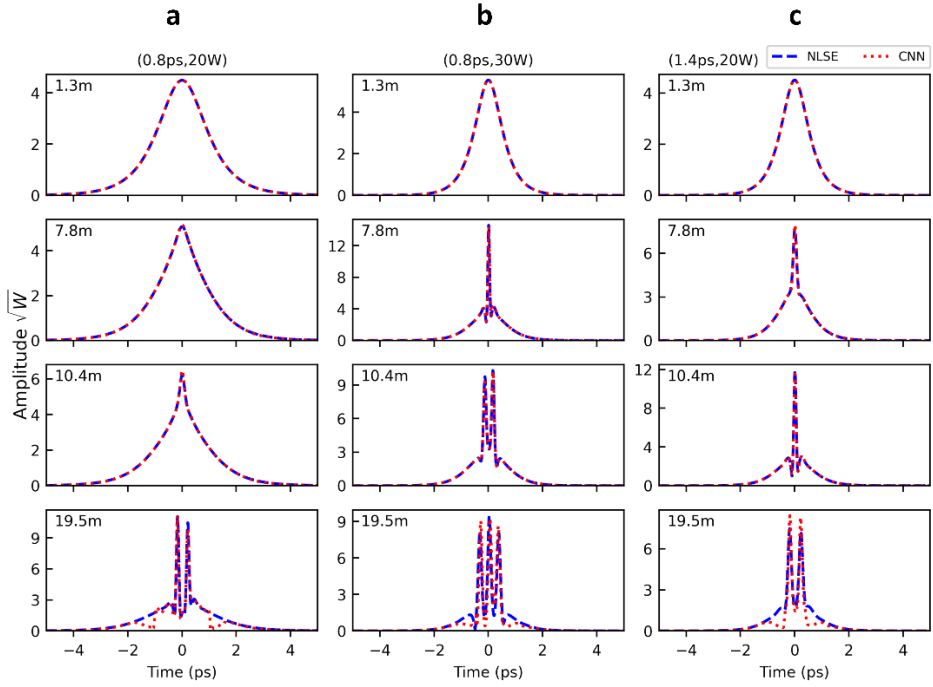

Fig. 6. Generalization capabilities validation of CNN with FSM for high-order soliton temporal dynamics. **a-c**, Pulse intensity profile at selected distances for different input conditions, including 0.8ps duration and 20W input peak power (a), 0.8ps duration and 30W input peak power (b), and 1.4ps duration and 20W input peak power (c). The distances presented are 1.3m, 7.8m, 10.4m, and 19.5m. Among them, 19.5m exceeds the maximum length in the training dataset (13m).

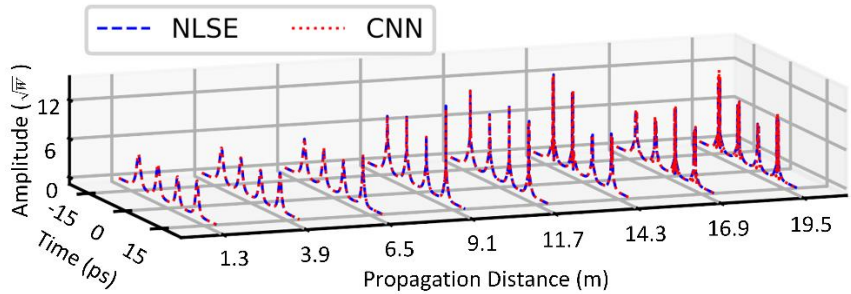

Fig. 7. Propagation of multiple pulses in single simulation. The pulse in the training dataset is only one. During the testing process, the input points number can be elongated to form multiple pulses with different conditions.

## Complexity

In order to compare the complexity of RNN and CNN, we theoretically calculate the total numbers of NN parameters and multiplication computations. The parameters number directly reflect the running memory and multiplication number is a common criterion to evaluate the complexity of an algorithm [27].

The CNN structure is described in detail above. The RNN structure we test is the same as that in [17], and also described above. Although both models can reduce the dimension of the input layer through the operation of truncating input pulse, to more fairly compare the complexity of the two structures, we do not truncate when calculating the number of parameters and the multiplications. Fig. 8a illustrates the number of parameters vs. the pulses grid points number of RNN and CNN. One can see that the parameter quantity of RNN increases quadratically with the number of input points since the dimension of the hidden layer is the same as the input layer. If the hidden dimension is fixed at a small level, the number of parameters can be reduced with a linear trend, but the accuracy will be reduced. The parameters of CNN remain unchanged with the increase of points, because the designed CNN structure is independent of the number of input pulses grid, as shown in Section 2. The parameters required by CNN are much less than RNN. Specifically, RNN needs about 14 billion parameters at the 2048 pulse grid points, while CNN remains constant at about 200 thousand. Namely, the required parameters of CNN are only 0.14% of RNN at the 2048 pulse grid points, which achieves a magnitude of decrease. The much smaller parameter count can greatly reduce the storage cost.

The number of multiplications vs. the grid points of RNN and CNN is illustrated in Fig. 8b. The results show the nearly linear increase of CNN and the quadratic increase of RNN. The trend with the pulse grid points is determined by the structure and dimension of NN. The multiplication numbers of CNN and RNN with 2048 pulse points are 400 million and 14 billion respectively. With the increase of grid points number, the complexity reduction benefited from the CNN structure becomes more obvious. The much fewer multiplication number of CNN demonstrates the better low computational complexity of CNN.

We also record the running time of the CNN, RNN, and NLSE under the same hardware and software conditions. The NLSE is implemented by the classical separate step Fourier method. The channel conditions setting is the same with SC generation. The codes of these three methods are run on the same server with two Intel Xeon Gold 6146R processers using Python. The model for each condition runs five times and get an average value as the time results. As shown in Fig. 9, the time increases with the grid points and the realization number on the central processing unit (CPU), and the CNN time is shorter than RNN and NLSE. For example, 10 realizations for different pulses with 2048 points take about 763s by NLSE, 46s by RNN, and 5.9s by CNN. The running time of CNN achieves a 94% reduction versus NLSE and an 87% reduction versus RNN. If NNs are run on the graphics processing unit (GPU), the time can be further reduced.

In addition, the input pulse can be truncated to accelerate the operation time as shown in Fig. 2. This is done dynamically and does not require retraining the model. In the HOS compression and 100 realizations case, the input pulse includes 1024 complex points initially and it may be truncated to 760 points with a 5% increase in RNMSE, and the time can be reduced to 13s from 19s. In SC generation, the input pulse includes 2048 complex points initially and it may be truncated to 970 points with a 0.2 % increase in RNMSE, and the time can be reduced to 18s from 50s with 100 realizations.

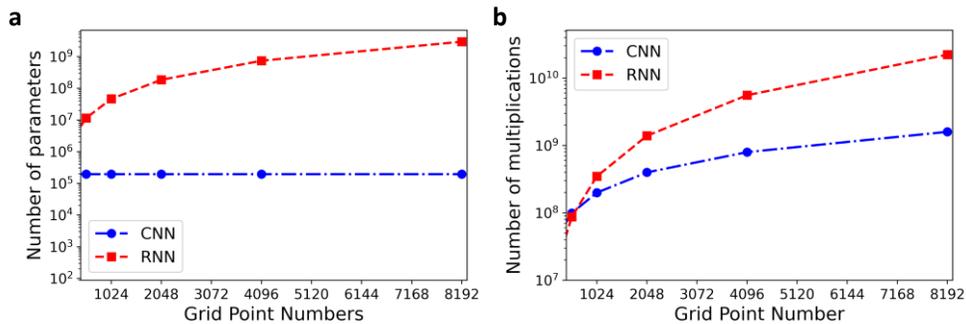

Fig. 8. The number of parameters (a) and multiplications (b) of CNN and RNN. The grid point number refers to the sampled pulse length input to the neural network. Note that the input pulses are not truncated for a fair comparison.

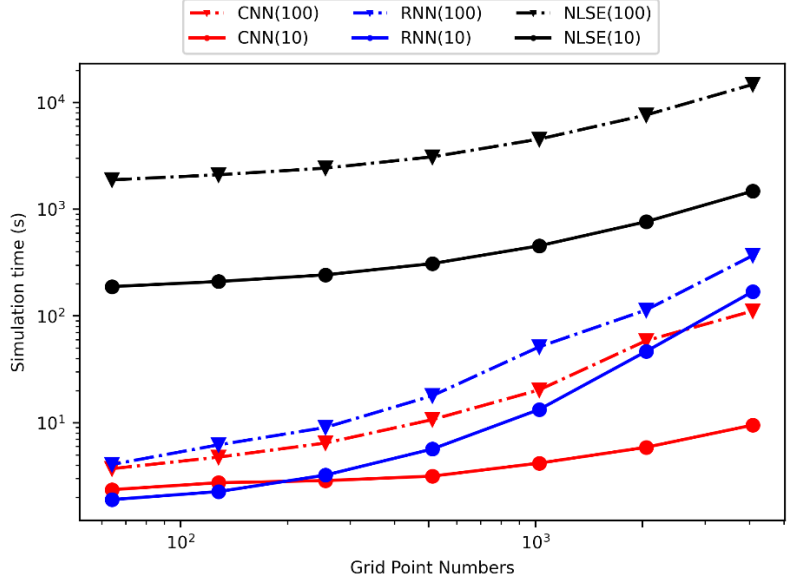

Fig. 9. The running time to compute evolution maps of supercontinuum using NLSE, RNN, and CNN with FSM on CPU. The transmission distance is 20cm, and the number of computed maps is 10 and 100.

The above results demonstrate that CNN with FSM can accurately, flexibly and rapidly predict the full-field ultrafast nonlinear dynamics. The intensity and phase information of the full-length input pulse in both temporal and spectral domains can be predicted. Compared with RNN, the CNN with FSM presents equal accuracy within the training transmission distances and higher accuracy beyond the training distances. Since the local convolution computation, the input pulse length can be flexibly shortened and elongated. This work provides a more flexible nonlinear dynamics prediction method. In addition, after the linear feature modeling by the traditional model-driven method, the residual nonlinear feature can be regarded as short-time dependent relations. Then the CNN structure with fewer parameters can be applied for local feature computation, which leads to a prediction method with a much shorter running time. Finally, the CNN with FSM can achieve high accuracy, robust generalization, and low complexity.

## 4. Conclusion

In conclusion, we propose a convolutional FSM for nonlinear dynamics prediction of the full-field ultrafast pulse evolution. Compared to the existing methods, our proposed method generalizes better and runs faster because of the designed small-scale CNN structure and the linear-nonlinear feature separation operation. The high-degree overlaps of intensity and phase profiles and low RNMSEs demonstrate the accuracy of our proposed method. The lower complexity results in an 87% reduction in computing time compared to the RNN in the SC generation case. We believe this FSM method with the computation of the local correlation will have positive impacts on future nonlinear physics researches attributed to the flexible, high-accuracy, and low-complexity design for nonlinear dynamics analysis and prediction.

**Funding.** This work is supported by the National Natural Science Foundation of China (62025503).